\documentclass[11pt,superscriptaddress,nofootinbib]{article}
\usepackage{cite}
\usepackage{mathptmx}
\usepackage{amsmath,amsfonts,amssymb,bbm}
\usepackage{lmodern}
\usepackage{simplewick}
\usepackage[small,bf,hang]{caption}
\usepackage[bookmarks=true,hyperfigures=true,hidelinks]{hyperref}
\usepackage{slashed}
\usepackage{latexsym,epsfig}
\usepackage{color}
\usepackage{mathtools}
\usepackage{nicefrac}
\usepackage{mathrsfs}

\def\hybrid{
        \topmargin -20pt
        \oddsidemargin 0pt
        \headheight 0pt \headsep 0pt
        \textwidth 6.25in 
        \textheight 9.5in 
        \marginparwidth .875in
        \parskip 5pt plus 1pt \jot = 1.5ex}

\hybrid

 \csname
@addtoreset\endcsname{equation}{section}

\def\moth{\mathsurround=0pt}
\newdimen\zo \zo=0pt

\def\tick{\leaders\hrule height 0.5ex depth 0pt \hskip 0.5pt}
\def\upboxfill{$\moth \setbox\zo\hbox{\tick}%
  \hskip 3pt\hbox to 0pt{$\tick$\hss}\hrulefill \hbox to 7.5pt{$\tick$\hss}$}

\def\dtick{\leaders\hrule height .34pt depth 0.5ex \hskip 0.5pt}
\def\downboxfill{$\moth \setbox\zo\hbox{\dtick}%
  \hskip 2pt\hbox to 0pt{$\dtick$\hss}\hrulefill \hbox to 2pt{$\dtick$\hss}$}

\def\al{\alpha}
\def\be{\beta}

\def\de{\delta}

\def\eps{\varepsilon}

\def\vp{\varphi}

\def\E{{\rm E}_{10}}
\def\KE{{{\rm K}(\E})}

\def\bfa{{\boldsymbol{{\tt a}}}}

\def\bbfa{\bar{{\boldsymbol{{\tt a}}}}}

\def\cI{{\cal I}}

\def\bi{{\bf a}}
\def\bj{{\bf b}}
\def\bk{{\bf c}}
\def\bl{{\bf d}}

\def\bE{{\bf 1}}
\def\bZ{{\bf 2}}
\def\bD{{\bf 3}}
\def\bV{{\bf 4}}

\def\bS{{\bf 6}}
\def\bA{{\bf 8}}

\def\bbE{{\bar{\bf 1}}}
\def\bbZ{{\bar{\bf 2}}}
\def\bbD{{\bar{\bf 3}}}
\def\bbV{{\bar{\bf 4}}}
\def\bbS{{\bar{\bf 6}}}

\def\bbi{{\bar{\bf a}}}
\def\bbj{{\bar{\bf b}}}
\def\bbk{{\bar{\bf c}}}
\def\bbl{{\bar{\bf d}}}

\def\mG{\mathfrak{G}}
\def\mN{\mathfrak{N}}

\def\eq{\!=\!}

\def\cI{{\mathcal I}}
\def\cJ{{\mathcal J}}

\def\cR{{\mathcal R}}
\def\cU{{\mathcal U}}

\def\cZ{{\mathcal Z}}
\def\11{{\mathbb 1}}

\def\ri{{\rm i}}
\def\rj{{\rm j}}

\def\rI{{\rm I}}

\def\beq{\begin{equation}}
\def\eeq{\end{equation}}
\def\bea{\begin{eqnarray}}
\def\eea{\end{eqnarray}}
\def\nn{\nonumber}

\def\ra{\rightarrow}
\def\Ra{\Rightarrow}

\def\ri{\text{i}}

\begin{document}

\begin{titlepage}
\begin{flushright}    
{\small $\,$}
\end{flushright}
\vskip 1cm
\centerline{\huge{\bf{ Standard Model Symmetries and $\KE$}}}
\vskip 1.5cm
\begin{center}
\begingroup\scshape\Large
Krzysztof A. Meissner$^{\,\diamond}$ and
Hermann Nicolai$^{\,\star}$
\endgroup
\end{center}
\centerline{{\Large ${\,}^\diamond$}\it {Institute of Theoretical Physics}}
\centerline {\it {University of Warsaw, Pasteura 5, 02-093 Warsaw, Poland}}
\vskip 0.5cm
\centerline{{\Large ${\,}^\star$}\it {Max-Planck-Institut f\"ur Gravitationsphysik (Albert-Einstein-Institut)}}
\centerline {\it {Am M\"{u}hlenberg 1, 14476 Potsdam, Germany}}
\vskip 1.5cm
\centerline{\bf {Abstract}}
\vskip .5cm
\noindent
We clarify and extend our earlier work  \cite{MN,MNPRL} 
where it was shown how to amend a scheme originally proposed by 
M. Gell-Mann to identify the three families of quarks and leptons 
of the Standard Model with the 48 spin-$\frac12$ fermions of 
$N\eq 8$ supergravity that remain after absorption of eight Goldstinos,
a scheme that in its original form is dynamically realized at the SU(3)$\,\times\,$U(1)
stationary point of gauged $N\eq 8$ supergravity.
We explain how to deform and enlarge this symmetry 
at the kinematical level to the full Standard Model
symmetry group SU(3)$_c\,\times\,$SU(2)$_w\,\times\,$U(1)$_Y$,  
with the correct charge and chiral assignments for all fermions.
The framework also leaves room for an extra U(1)$_{B-L}$ symmetry. 
This symmetry enhancement is achieved by embedding the Standard Model  
symmetries into (a quotient group of)  $\KE$, the `maximal compact subgroup' of 
the maximal rank hyperbolic Kac-Moody symmetry $\E$,  
and an infinite prolongation of the SU(8) $R$-symmetry of
$N=8$ supergravity. This scheme, which is also supposed to
encompass quantum gravity, cannot be
realized within the framework of space-time based (quantum) 
field theory, but requires space-time and related geometrical concepts to 
be `emergent'.  We critically review the main hypotheses underlying this construction. 
\vfill
\end{titlepage}


\section{Introduction}
The outstanding challenge in any attempt at unifying the fundamental
interactions is to come up with an explanation of the Standard Model (SM) 
symmetries and degrees of freedom, and to answer the question why its 
fermions come in three families (generations) of quarks and leptons, 
including right-chiral neutrinos. The conventional strategy for achieving this goal 
is based on enlarging SM symmetries, by promoting the
SM gauge group to a GUT group, and/or by embedding the SM into a 
supersymmetric extension, ultimately in the framework of superstring 
theory so as to incorporate quantum gravity. The SM is then 
supposed to emerge from such a more unified theory in a cascade of 
symmetry breakings, step by step descending from the unification scale
to the electroweak scale. Unfortunately, this strategy has so far not yielded
a  compelling explanation of why the SM is the way it is, and not
otherwise, and in particular cannot explain the three-fold replication 
of fermion generations. Rather, the number of generations is a parameter 
that can be freely dialled, for instance by choosing an appropriate 
Calabi-Yau manifold in superstring compactifications. Furthermore, and
especially in connection with low energy supersymmetry, these
ans\"atze generically come with numerous new fermionic and bosonic  
degrees of freedom, for which there is no evidence whatsoever 
after more than 15 years of LHC operation. The staggering vacuum 
degeneracy of superstring theory has even led to suggestions
that a multiverse interpretation is the only plausible road
towards explaining the world we live in.

Following up on our earlier work \cite{MN,MNPRL} we here describe 
a possible alternative path towards such an explanation, which 
tries to make do with the known three generations of quarks 
and leptons, and which aims for a more or less direct explanation from
a Planck scale theory. This more rigid framework also means
that this proposal is eminently falsifiable. Our approach is based on
a colossal enlargement of symmetries beyond the SM gauge group
SU(3)$_c\,\times\,$SU(2)$_w\,\times\,$U(1)$_Y$, and transcends
established field theoretic concepts by invoking the concept of an
emerging space-time, and thus quantum gravity,  in an essential way. It thus
differs substantially from currently popular paradigms to derive the structure
of SM fermion multiplets from a Planck scale unified theory of quantum 
gravity. It relies on the key conjecture of \cite{DHN}, according to 
which the `maximally extended' hyperbolic Kac-Moody $\E$ algebra is a central
part of the symmetries underlying quantum gravity and the unification  
of fundamental interactions~\footnote{In \cite{West} a conceptually 
 very different scheme has been proposed which is based on the indefinite
 (but non-hyperbolic) Kac-Moody algebra E$_{11}$.}. This conjecture 
 posits that this symmetry reveals itself only in a `near singularity limit', 
 which requires a framework beyond space-time based (quantum)
 field theory. It is itself based on a BKL-type analysis of  cosmological 
 singularities, where the causal decoupling of spatial points entails 
an effective dimensional reduction to one (time) dimension \cite{BKL}.
In this pre-geometric regime, the spatial dependence of the fields, and
with it, the usual field-theoretic structures and space-time symmetries,
are supposed to `de-emerge' near the singularity, in the sense that the
spatial dependence gets `spread' over the Lie algebra of $\E$. 
Of even greater importance for the present proposal is the `maximal compact'
involutory subalgebra $\KE \subset \E$ which governs the fermionic 
sector \cite{DKN,dBHP}, and which has been proposed as a generalized
`$R$-symmetry', as it is an infinite-dimensional prolongation of the 
$R$-symmetries SU(8) $\equiv$ K(E$_7$) and SO(16) $\equiv$ K(E$_8$)
of maximal supergravities.

The second central ingredient of the present work is an old proposal of 
Gell-Mann's \cite{GellMann} to identify the 48 quarks and leptons of the 
SM (including right chiral neutrinos) with the 48 spin-$\frac12$  fermions 
that remain after complete breaking of $N\eq 8$ supersymmetry and 
absorption of eight Goldstinos. That proposal was originally made in connection 
with maximal $N\eq 8$ supergravity \cite{CJ,dWN}, but fell short  
of fully explaining the SM charge assignments, without any option
of rectifying the mismatches within $N\eq 8$ supergravity.
One particular mismatch concerned the electric charges of the
quarks and leptons, which differ systematically by a `spurion charge' $\pm\frac16$
from the ones predicted by supergravity, see (\ref{Fermions1}) below.   
However, as shown in \cite{MN}, this mismatch of electric charges can 
be corrected by introducing a new vector-like generator $\cJ$
(see (\ref{cI}) below) which is not contained in the SU(8) $R$-symmetry 
group of $N\eq 8$ supergravity. In subsequent work \cite{KN} 
it was furthermore shown that this new generator can be realized within 
$\KE$,  thus confirming the need for new ingredients {\em beyond} 
$N\!=\!8$ supergravity. Remarkably, this embedding involves
{\em all} simple roots of $\E$ \cite{KN}, hence there
is no way to make the argument work with the affine involutory
group K(E$_9)\subset  \KE$ or its finite-dimensional subgroups.
In other words, there is no way in this set-up 
to explain the SM charges of the fermions without `going hyperbolic'.
Here we take this construction yet one step further by exploring how 
the full SU(3)$_c\, \times$ SU(2)$_w \, \times$ U(1)$_Y$ and U(1)$_{B-L}$ 
symmetries of the SM can be kinematically incorporated into $\KE$.
In addition there is a family symmetry SU(3)$_f$ which is, however, 
necessarily broken in the presence of electroweak symmetries in the SM -- 
in agreement with the real world where any putative family symmetry is 
broken by the Yukawa couplings.

Being restricted to the fermions, our construction relies in an essential 
way on the fact that $\KE$ possesses unfaithful {\em finite-dimensional} 
fermionic representations corresponding to the fermions of maximal 
supergravity {\em at a given spatial point} on the initial 
hypersurface \cite{DKN,dBHP}, in accord with the BKL conjecture. 
It is a remarkable fact here that even though accompanied 
by  a vast enlargement of the $R$-symmetry, our proposal makes 
do in a first step with a finite number of fermionic degrees of freedom, 
whereas more conventional schemes would require correspondingly 
larger multiplets for larger groups. It is this feature which makes 
our considerations possible in the first place. Nevertheless, it is
clear that the completion of the program beyond the `point-like'
action of the SM symmetries on the fermions will eventually require
infinite-dimensional faithful representations.
By contrast, in the bosonic sector there is
no way of consistently truncating the action of the $\E$ symmetry
to the lowest levels corresponding to the field values and their first
order spatial gradients at a given spatial point. Rather this action 
on the bosons mixes all levels in an $\E$ level decomposition (see {\em e.g.}
\cite{NF} to get a flavor of the exponential growth of such decompositions).
This suggests that the identification of the proper physical bosonic 
degrees of freedom in this framework is more complicated.
This is the main reason why not more can be said 
about the bosonic SM degrees of freedom at this point. 

Even though the construction is purely kinematical (a bit like
considerations of strong interaction physics in the early 1960ies)
one can still try to exploit the `proximity' of $N\eq 8$ supergravity,
for which at least the classical dynamics is better understood. 
The very existence of a stable SU(3)$\,\times\,$U(1) stationary 
point of the scalar field potential is an important fact \cite{Warner}, 
as is the fact that it gets so close to a realistic scenario, given the
existing evidence from particle physics and astrophysics that 
there are only three families of quarks and leptons in Nature.
Independently of any remaining open issues, our scheme thus makes
three definite predictions for `Beyond the SM' physics, which are such 
that the proposal stands or falls with them:
\begin{enumerate}
\item There are no fundamental spin-$\frac12$ fermions beyond
          the observed three families of quarks and leptons of the SM,
          including right-chiral neutrinos. In particular, the number three
          of families is a definite and unalterable prediction.
\item  The only new fermions are spin-$\frac32$ supermassive gravitinos, 
          which carry SM charges, with only vector-like interactions, whence their
          natural mass scale is the Planck scale. Nevertheless, being subject
          to SM interactions, these particles can in principle be detected
          in upcoming underground experiments like JUNO or DUNE \cite{KLMN}.
\item  The apparent incompatibility of the rectified charge assignments in (\ref{Fermions2})
          with space-time supersymmetry implies either the existence of a novel 
          supersymmetry breaking mechanism at the Planck scale, or the complete absence of 
          space-time supersymmetry at any energy scale.         
\end{enumerate}

The first point means that our proposal can be directly falsified by the 
discovery of any new fundamental spin-$\frac12$ fermion, such as a fourth family
of quarks and leptons, or new sterile neutrinos beyond the observed three, 
or any new fermion predicted by low energy supersymmetry (gaugino,
Higgsino, and the like). We note that there are only few other `Beyond the SM'
scenarios that try to make do with the observed spin-$\frac12$ fermions (one 
prominent example is the $\nu$MSM model of \cite{Shap}), but usually these do not 
address the question of quantum gravity, whereas $\E$ from the outset treats 
gravitational degrees of freedom on a par with non-gravitational (matter) 
degrees of freedom. These models often incorporate enlarged scalar sectors 
(`Higgs portals'), but this is an issue on which we have nothing new to 
say here, for the reasons stated above.

The second item is the one that offers the best options for experimentally 
validating our proposal. While it seems clear that supermassive gravitinos 
cannot be produced in any conceivable collider experiment, \cite{KLMN}
explains in detail how to search for them with upcoming underground 
detectors, even if those experiments are mainly designed for other purposes
(neutrino physics, in particular).  Detecting traces of a supermassive 
gravitino would be an encouraging signal that the present ideas are on the 
right track!  Apart from a direct detection, supermassive
gravitinos can manifest themselves  in other ways, and more indirectly.
Most importantly, they are new Dark Matter candidates \cite{MN1}, of an 
entirely different kind from those considered so far in the literature (see {\em e.g.} 
\cite{CSZ} for a comprehensive review). Supermassive gravitinos might
also help in addressing two important open problems in astrophysics:
strongly interacting color triplet gravitinos might play a role in explaining 
the origin of ultrahigh energy cosmic rays via mutual annihilation of gravitinos
in the `skin' of neutron stars \cite{MN2}, and supermassive gravitinos can 
serve as seeds for primordial black holes via the formation and gravitational
collapse of `gravitino lumps' in the early radiation period \cite{MN3}.

Item three could provide a way out of the dilemma posed by the no-show 
of supersymmetry in particle physics. In spite of the undeniably attractive 
features of supersymmetric quantum field theories, there is so far no 
evidence at all that Nature wants to make use of it. Furthermore, finding 
a compelling mechanism to break supersymmetry has been a long 
standing problem which remains unsolved. The present construction points 
to a way of eliminating supersymmetry in the effective space-time
field theory, even though ($N\eq 2$) supersymmetry continues to play 
an important role `behind the scene', in that it fixes the fermion
multiplets and the charges in (\ref{Gravitino1}), (\ref{Goldstino}) and 
(\ref{Fermions1}) below. Nevertheless, there are
indications that space-time supersymmetry and $\E$ cannot
go together because the local supersymmetry algebra appears to be
incompatible with the presence of imaginary roots \cite{CKN}. A further 
potential incompatibility with space-time supersymmetry is the fact that, in the cosmological 
context, the $\E$ model does not appear to support the AdS-type solutions 
strongly preferred by supersymmetry \cite{E10cosmo}.

To conclude we emphasize that key questions remain
wide open. Essential elements are still missing before one can truly understand 
the emergence of (quantum) field theoretic concepts from a purely algebraic construct. 
In particular, explaining the emergence of a space-time dependence and
understanding the role of higher level modes in the M theory proposal of \cite{DHN}
remains an outstanding challenge, as well as the question whether the
$\E$ framework needs to be supplemented (as suggested by recent work
on E$_{11}$ exceptional field theory \cite{Bossard}).
Altogether, the present construction can therefore only 
be a very first step in an ongoing effort to relate the abstract concepts
developed in connection with $\E$ and $\KE$ to SM physics.

The organization of this paper is as follows. In section~2 we review
the SU(3)$\,\times\,$U(1) breaking of $N\eq 8$ supergravity, and the
resulting decomposition of the fermions of the $N\eq 8$ supermultiplet 
which turns out to be uniquely fixed by $N\eq 2$ supersymmetry \cite{NW}.
In section~3 we explain how to realize all SM symmetries on these
fermions by extending the $R$-symmetry SU(8). Section~4 reviews some
basic features of $\KE$ (see \cite{Vigano} for a much more detailed
review), and explains how the SM symmetries can be embedded 
into $\KE$ via the relevant quotient subgroups. Finally, in section~5
we recall some basic features of maximal gauged supergravities which
possess a SU(3)$\,\times\,$U(1) vacuum, and argue that these vacua 
are distinguished. A short summary of the open problems concludes the paper.

\section{Fermions and the SU(3)$\,\times\,$U(1) vacuum of $N\eq 8$ supergravity}

As is well known \cite{CJ,dWN}, the fermionic part of the $N\eq 8$ supermultiplet 
comprises eight gravitino vector spinors $\psi^i_{a \alpha}$  transforming 
in the $\bA$ of SU(8), and a tri-spinor of spin-$\frac12$ fermions $\chi^{ijk}_\alpha$ 
transforming in the $\bf{56}$ of SU(8), where  $\chi^{ijk}_\alpha$ 
is fully antisymmetric in the SU(8) indices $i,j,k$ (we restrict attention
to the {\em spatial} components of the vector spinor, adopting a temporal
gauge for the gravitino as in \cite{DKN}). We adopt standard SL(2,$\mathbb{C}$)
spinor conventions, where complex conjugation raises (or lowers) indices, such that
$\psi^i_{a\alpha} = \big( \bar\psi_{\dot\alpha\,i}\big)^*$ 
and  $\chi^{ijk}_\alpha = \big( \bar\chi_{\dot \alpha ijk} \big)^*$, and
where upper  (lower)  position of the SU(8) indices indicates positive 
(negative) chirality. Hence  the chiral SU(8) transformations act as
\bea\label{Uchi}
\chi^{ijk}_\alpha \; & \ra& \;  U^i{}_l U^j{}_m U^k{}_n \chi^{lmn}_\alpha \quad , \nn\\[2mm]
\bar\chi_{\dot\alpha\, ijk}  \; &\ra&  \;  U_i{}^l U_j{}^m U_k{}^n \bar\chi_{\dot\alpha\, lmn}
\eea
with $U \in$ SU(8), $U_i{}^j \equiv \big( U^i{}_j \big)^*$ and $U^i{}_k U_j{}^k = \delta^i_j$;
for the maximal vector-like subgroup SO(8) $\subset$ SU(8) 
$U$ and $U^*$ coincide. In the remainder, we will only use
left-chiral fields,  {\em i.e.} describe right-chiral particles 
by left-chiral anti-particles, in accordance with the $CPT$-theorem,
and will henceforth suppress SL(2,$\mathbb{C}$) indices. 

As shown long ago by N.P.~Warner, SO(8) gauged maximal supergravity admits
a stationary point with residual SU(3)$\,\times\,$U(1) symmetry \cite{Warner}. 
This  group is contained in the vector-like gauge group SO(8) via the 
standard embedding U(4)$\,\subset$ SO(8). For this decomposition, we use boldface 
indices $\bE,...,\bV$,  and barred indices $\bbE,...,\bbV$, with
$v^\bE\equiv 2^{-1/2} (v^1 + iv^2), 
\cdots ,v^\bV \equiv 2^{-1/2}(v^7 + iv^8)$, 
$v^\bbE \equiv 2^{-1/2}(v^1 - iv^2), \cdots, v^\bbV \equiv 2^{-1/2}(v^7 - iv^8)$.
With this notation we have, for instance,
\bea\label{vp}
\vp^{\bE\bZ\bbV} &\equiv& \frac1{2\sqrt{2}} \Big(\chi^{137} + i\chi^{237} + i\chi^{147} - i\chi^{138} \nn\\
          && \qquad -\chi^{247} + \chi^{238} + \chi^{148} + i \chi^{248}\Big) \, ,
\eea          
for the tri-spinor, where we use the letter $\vp$ instead 
of $\chi$ when referring to this basis for the 56 spin-$\frac12$ fermions.
Note that the spinor obtained by interchanging barred and
unbarred indices
\bea\label{vpb}
\vp^{\bbE\bbZ\bV} &\equiv& \frac1{2\sqrt{2}} \Big(\chi^{137} - i\chi^{237} - i\chi^{147} + i\chi^{138} \nn\\
          && \qquad -\chi^{247} + \chi^{238} + \chi^{148} - i \chi^{248}\Big) \, ,
\eea    
is {\em not}  the complex conjugate of $\vp^{\bE\bZ\bbV}$ because the 
original tri-spinor components $\chi^{ijk}$ are themselves complex (the
fact that (\ref{vp}) and (\ref{vpb}) represent {\em independent} degrees of
freedom would be completely obvious if we reverted to more standard
conventions with left- and right-chiral spinors).
Next we single out the first three indices with labels 
$\bi ,\bj, \dots= \bE,\bZ,\bD$, on which the SU(3) subgroup acts. 
Then there is a two-parameter family of U(1) subgroups
of SO(8) with generators
\begin{equation} \label{U(1)}
 \cZ_{\al,\be} \,= \, \begin{pmatrix}
                0 & - \al  & 0 & 0 & 0 & 0 & 0 & 0 \\
                \al  & 0 & 0 & 0 & 0 & 0 & 0 & 0 \\
                0 & 0 & 0 & -\al & 0 & 0 & 0 & 0 \\
                0 & 0 & \al & 0 & 0 & 0 & 0 & 0 \\
                0 & 0 & 0 & 0 & 0 & - \al & 0 & 0 \\
                0 & 0 & 0 & 0 & \al & 0 & 0 & 0 \\
                0 & 0 & 0 & 0 & 0 & 0 & 0 & -\beta\\
                0 & 0 & 0 & 0 & 0 & 0 & \beta & 0
                \end{pmatrix},
\end{equation}
This matrix commutes with U(4) $\subset$ SO(8), and thus defines 
an SU(3)$\,\times\,$U(1) subgroup of SO(8) for every choice of $\al$ and $\be$.
In order to match the $N\!=\!8$ fermions with those of the SM, 
the requisite choice is \cite{GellMann}
\beq\label{albe}
\al = \frac16 \;\; , \quad\be = \frac12 \, .
\eeq
These values are also obtained for the SU(3)$\,\times\,$U(1) 
stationary point of the scalar potential of $N\eq 8$ supergravity, 
and in fact uniquely fixed (up to scaling) by the structure of the relevant 
$N\!=\!2$ AdS supermultiplets \cite{NW}.  The special matrix
\beq
\cI \,\equiv \, \cZ_{1,1}
\eeq
serves as an imaginary unit for SU(3)$\,\times\,$U(1), 
in the sense that
$(\cI \circ v)^\bi = i v^\bi \,, \, (\cI \circ v)^\bbi = -i v^\bbi$ and  
$(\cI \circ v)^\bV = i v^\bV \,, \, (\cI \circ v)^\bbV = -i v^\bbV$.
Because $\cI$ is a real matrix, SU(3)$\,\times\,$U(1) is realized
as a real group on the tri-spinor $\chi^{ijk}$.

With the choice (\ref{albe}) we obtain the following SU(3)$\,\times\,$U(1) 
assignments for the gravitinos 
\bea\label{Gravitino1}
\psi_a^\bi \;&\in&\;  (\bD\,,\, {\small\frac16)}\; , \quad \psi_a^\bbi \;\;\in \;\;(\bbD\,,\, - \frac16)\; ,  \nn\\[2mm]
\psi_a^\bV \; &\in&\; (\bE \,,\,\frac12) \; , \quad \psi_a^\bbV \;\;\in\;\; (\bE \,,\, - \frac12)
\eea
The 56 spin-$\frac12$ fermions are split into 6\,+\,2 Goldstinos
\bea\label{Goldstino}
\vp^\bi &\equiv& \vp^{\bi\bV\bbV} \,\in\, (\bD\,,\, \frac16) \; , \;\;
\vp^\bbi \equiv \vp^{\bbi\bV\bbV} \,\in \, (\bbD\,,\, - \frac16) \;\; , \nn\\[2mm]
\vp^\bV &\equiv& \vp^{\bi\bj\bk} \in (\bE\,,\, \frac12) \;,\;\;
\vp^\bbV \equiv \vp^{\bbi\bbj\bbk} \,\in \,(\bE\,,\, -\frac12)
\eea  
and the remaining 48 spin-$\frac12$ fermions \cite{GellMann,NW}

\bea\label{Fermions1}
\vp^{\bi\bj\bV} &\in& (\bbD\,,\, \frac56) \; , \quad \vp^{\bi\bj\bbV} \,\in\, (\bbD\,,\, - \frac16)\,,\quad 
\\[1mm]
\vp^{\bbi\bbj\bV} &\in& (\bD\,,\, \frac16) \; , \quad \vp^{\bbi\bbj\bbV}\,\in\, (\bD\,,\, - \frac56) \nn
\\[1mm]
\vp^{\bi\bj\bbk} \,&\in& \, (\bD,\frac16) \oplus (\bbS,\frac16) \;,\quad
\vp^{\bbi\bbj\bk} \,\in \, (\bbD, - \frac16) \oplus (\bS\,,\,-\frac16) \;\;
\nn\\[1mm]
\vp^{\bi\bbj\bV} \,&\in&\, ({\bf 8}\,,\,\frac12) \oplus (\bE\, ,\,\frac12) \;,\quad
\vp^{\bi\bbj\bbV} \,\in\,  ({\bf 8}\,,\, - \frac12) \oplus (\bE\,,\, -\frac12)  \nn
\eea
In order to identify the SU(3) singlet Goldstinos, we must in principle allow 
for a mixing between the representations $(\bE,\pm\frac12)$ which
occur twice in the decomposition (such mixing does occur for the mass 
eigenstates at the SU(3)$\,\times\,$U(1) stationary point \cite{NW}).
However, the precise form of this mixing depends on the dynamics 
which at this point is not known. As we will see below, the above 
choice is singled out if one wants to disentangle the two SU(3)'s
in Gell-Mann's scheme and implement a proper action of the
color group SU(3)$_c$ on the fermions. It is therefore an essential 
assumption that we can bring the fermions  to the form with the 
labeling exhibited above, possibly by means of a redefinition of the 
spin-$\frac12$ fermions within the enlarged $R$-symmetry $\KE$.

Now, as first pointed out in \cite{GellMann}, with the above choice 
of Goldstinos the three generations of SM fermions can be matched 
with the remaining 48 spin-$\frac12$ fermions from (\ref{Fermions1}), 
provided one identifies the  supergravity SU(3) with the diagonal 
subgroup of color SU(3)$_c$ and a hypothetical family symmetry SU(3)$_f$, 
{\it viz.}
\beq\label{diag}
{\rm SU}(3) \equiv \Big[ {\rm SU}(3)_c \times {\rm SU}(3)_f \Big]_{\rm diag}
\eeq
and furthermore allows for a spurion charge shift $\pm \frac16$ to correct 
the U(1)charges in (\ref{Fermions1}) so as to recover the known electric charges 
of quarks and leptons. Nevertheless, the above charge and representation
assignments illustrate that supergravity can take us only this far, but no further: there 
is no way {\em within} maximal supergravity to rectify the U(1) charge mismatches,
nor to undo the descent to the diagonal SU(3) subgroup in (\ref{diag}).

In addition to its SU(3)$\,\times\,$U(1) gauge symmetry this 
stationary point of the supergravity potential exhibits a 
residual $N\eq 2$ local supersymmetry, for which the relevant
AdS supermultiplets have been worked out in \cite{NW}. Therefore
we have six massive gravitinos of Planck scale mass (as well as several
massive $N=2$ supermultiplets). For the present proposal
to work the remaining two supersymmetries must then be 
broken by a novel mechanism, as supersymmetry appears to be
incompatible with the adjustments required to make the charges 
agree with those of the SM fermions. As a result we may end up with 
two different mass scales for the gravitinos, with the SU(3) triplet
gravitinos of Planck mass, and the remaining two gravitinos 
acquiring masses (somewhat) below the Planck scale. As we argue
below, only the introduction of chiral interactions can keep all the
remaining fermions (almost) massless in comparison with the Planck scale.

\section{Implementing SM symmetries on the fermions}

In this section we show how to realize all SM symmetries
on the fermions of the $N\eq 8$ supermultiplet at the kinematical level. 
These arguments go beyond the results of \cite{MNPRL} in that we provide 
the proper identification of the fermions for the SM symmetries to act on,
as well as explicit formulae for the electroweak hypercharge in 
(\ref{T3}) and the U$_{B-L}$-charge in (\ref{BmL}) below. In the following
section we will recall how to embed this picture into the
M theory conjecture of \cite{DHN}, with a huge enlargement of
symmetries beyond supergravity. In accordance with this conjecture
the full symmetry manifests itself only in a `near singularity limit'
which effectively amounts to a reduction of maximal supergravity 
to one (time) dimension (for which the appearance of $\E$ was
originally conjectured in \cite{BJ}). Consequently, the fermionic variables 
below are to be understood as the values of the fermionic fields 
at a given spatial point, {\em i.e.}
\bea\label{spatial}
\chi^{ijk} \,&\equiv&\,  \chi^{ijk}(t)  \,\equiv\, \chi^{ijk}(t, {\bf x}_0) \;\;, \nn\\[2mm]
\psi^i_a \, &\equiv&\, \psi^i_a (t)  \,\equiv\, \psi^i_a (t, {\bf x}_0) 
\eea
where ${\bf x}_0$ is the chosen fixed spatial point. The incorporation of the 
full spatial dependence and the extension of the point-wise group actions
to space-time dependent gauge transformations remain open problems
that can presumably only be resolved with a much better understanding 
of $\E$ and $\KE$.

The requisite symmetry enhancement requires extensions 
beyond the SU(8) $R$-symmetry of $N\eq 8$ supergravity which we now
spell out. To undo the diagonal action and to enable a separate 
(and commuting) action of the two SU(3)'s we take the 48 non-Goldstino 
components of the tri-spinor apart as follows, 
already with the rectified charge assignments to be explained below:

\begin{align}
\label{Fermions2}
&{\mbox{Tri-spinor}}& \
\, \, {\rm SU}(3)_c&\,\times\,{\rm SU}(3)_f
& \ Q_{em} & \qquad\qquad  \,Y & Q_{B-L}\nn\\[3mm]
&\vp^{\bi|\bbj\bV}\; & \
\, &\, \bD_c \times \bbD_f   
& \ \frac23 & \qquad\qquad  \,\frac16& \frac13
 \nn\\
 &\vp^{\bbi|\bj\bbV}\; &  \
 \, &\, \bbD_c \times \bD_f  
 & \ - \frac23  & 
\quad\qquad  -\frac23&\, -\frac13 \nn\\
&\vp^{\bi|\bbj\bbk}\; & \
\, & \, \bD_c \times \bD_f  
&
\  - \frac13 & \qquad\qquad   \,\frac16 &\, \frac13 \nn\\
&\vp^{\bbi|\bj\bk}\;  & \
\,& \, \bbD_c \times \bbD_f  
& \ \frac13  & 
\qquad\qquad \,  \frac13&\, -\frac13 \nn\\
&\vp^{\bbV|\bi\bj}\;  & \
 \, & \, \bE_c \times \bbD_f    
 & \  0 &
\quad\qquad  - \frac12 &\, -1\nn\\
&\vp^{\bV|\bbi\bbj}\; & \
\, & \, \bE_c \times \bD_f   
& \  0 &
\qquad\qquad  \, 0  &\, 1\nn\\
&\vp^{\bbV|\bbi\bbj}\;&  \
\, & \, \bE_c \times \bD_f   
& \ - 1 &
\quad\qquad   - \frac12&\, -1   \nn\\
&\vp^{\bV|\bi\bj}\; & \
\,  & \, \bE_c \times \bbD_f  
& \ 1  & \qquad\qquad \,   1&\, 1  
\end{align}
Here we have introduced a `$|$' as a mnemonic device to indicate that this index ordering
is now to be kept fixed. As we explain below, the 
above split  allows us to disentangle the actions of the two SU(3)'s, and furthermore
requires that the two SU(3) singlet Goldstinos in (\ref{Goldstino}) 
do not mix with the diagonal SU(3) singlets from $\bD\times\bbD$
in (\ref{Fermions2}). The implementation of larger symmetries thus 
would not work without  the prior removal of all Goldstinos, and 
more specifically the SU(3) singlet Goldstinos in (\ref{Goldstino}).  
Let us now explain how the symmetry enhancement works, and
in particular the charge assignments in the last three columns 
of (\ref{Fermions2}). We start with the electric charge assignments
in the third column of (\ref{Fermions2}), because this is 
what triggers all further steps.

\noindent{\bf Electric charges and U(1)$_{em}\,$:}
As was shown in \cite{MN} the 56-by-56 matrix operator that gives 
the correct values of electric charges of quarks and leptons  
listed in the third column is
\beq\label{electric}
iQ_{em} \,=\,   \frac16 \Big(\cZ_{1,3} \otimes {\bf 1} \otimes {\bf 1} \, + \,  
{\bf 1} \otimes \cZ_{1,3} \otimes {\bf 1} 
               \, + \,  {\bf 1} \otimes {\bf 1} \otimes \cZ_{1,3} \,+\, \cJ \Big)
\eeq
where
\beq\label{cI}
\cJ := \frac12 \Big( \cI \otimes {\bf 1} \otimes {\bf 1} \, + \,  {\bf 1} \otimes \cI \otimes {\bf 1} \, + \,
        {\bf 1} \otimes {\bf 1} \otimes \cI  \, + \,  \cI \otimes \cI \otimes \cI \Big)
        \quad \Ra  \;\; \cJ^2 = -1
\eeq
The first three terms on the r.h.s. of (\ref{electric}) correspond to the proper 
action of the U(1) subalgebra of SO(8) on the $\bf{56}$ tri-spinor 
$\chi^{ijk}$ via the usual co-product, giving the U(1) charges in
(\ref{Fermions1}) with the choice (\ref{albe}) 
dictated by supergravity. However, the last term with $\cJ$ 
requires an extension beyond the SU(8) $R$-symmetry (\ref{Uchi}). 
The operator $\cJ$ gives $(+i)$ on $\vp$'s with no or only one barred 
index, and $(-i)$ on $\vp$'s with two or three barred indices. 
Importantly, the operator  $\cI\otimes \cI \otimes \cI$, and hence 
the generator $\cJ$ are {\em not} elements of SU(8),
but still vector-like. The U(1)$_{em}$ defined by (\ref{electric}) still 
commutes with the original SU(3)$\,\times\,$U(1), which is thus 
preserved in a deformed form, but outside the original SO(8).
However, in \cite{KN} it was shown that the action of $\cJ$ on
the fermions can be reproduced by an appropriately chosen
element of $\KE$, and thus, by the unfaithfulness of the 
representation, by infinitely many elements of $\KE$
(see the following section). The
embedding into $\KE$ also shows that the action of $\cJ$
can be extended to the gravitinos, with $\cI\otimes\cI\otimes \cI$ 
replaced by $\cI$, so that\footnote{When acting on the gravitino
 the first three terms on the r.h.s. of (\ref{cI}) are simply replaced by 
 $\frac12 \cI$, corresponding to the proper action of SO(8) on the gravitino.}
\beq\label{cJgrav}
\cJ\circ \psi \,=\,  \cI \circ \psi \, . 
\eeq
Therefore the action of $\frac16 \cJ$ adds $\pm\frac16$ to the U(1) charges 
of the gravitinos in (\ref{Gravitino1}). Consequently, with this shift, $\psi_\mu^\bi$ has electric 
charge $\frac13$, and $\psi_\mu^\bV$ has electric charge $\frac23$, consistent 
with the charge assignments of the Goldstinos (\ref{Goldstino}) after
the spurion shift;  $\psi_a^\bbi$ and $\psi_a^\bbV$ 
have the opposite electric charges.

Having admitted $\cI\otimes\cI\otimes\cI$ as an extra symmetry
generator, we must include all generators obtained by commuting it with
the SU(8) from (\ref{Uchi}). A brute force computation based on repeated
commutation of $\cJ$ with the SU(8) generators yields the group SU(56) 
acting on the 56 spin-$\frac12$ fermions \cite{Vigano}. 
However, there is a much quicker argument to see this.  
The enlarged symmetry still acts irreducibly and unitarily on the 
original 56 fermions, thus {\em without} increasing the 
representation $\bf{56}$ of SU(8) to a larger one. Hence, the resulting 
group must also have a 56-dimensional irreducible representation. 
This leaves SU(56) or E$_7(\mathbb{C})$ as the only options. 
Because E$_7(\mathbb{C})$ is not unitary, the enlarged group can
only be SU(56); when restricting to commutators with the real group
SO(8), we get SO(56) instead. After the removal of eight Goldstinos 
this effectively leaves us with a chiral SU(48) symmetry acting on
the 48 spin-$\frac12$ fermions (\ref{Fermions2}), with SO(48)
as the maximal vector-like subgroup.

\noindent{\bf Disentangling SU(3)$_c\,\times\,$SU(3)$_f\,$:}
The SO(48) group generated by the inclusion of the new generator
$\cI\otimes\cI\otimes\cI$ contains SU(3)$_c\,\times\,$SU(3)$_f$ as 
a vector-like subgroup, and thus allows to disentangle the two SU(3)'s 
underlying  Gell-Mann's scheme. More specifically, the color SU(3)$_c$ 
acts on (\ref{Fermions2}) only on the first index via the matrix 
\beq\label{Uc}
\cU_c \,=\, U_c\otimes\bE\otimes \bE \qquad\quad
\mbox{for $U_c\in$ SU(3)$_c$}\;\;, 
\eeq
while the family SU(3)$_f$ acts only on the second and third indices by
\beq\label{Uf}
\cU_f \,=\, \bE\otimes U_f \otimes U_f  \qquad\quad
\mbox{for $U_f\in$ SU(3)$_f$} \;\;, 
\eeq
where the $U$'s act trivially on the indices $\bV$ and $\bbV$.  This action 
is unitary, but obviously not compatible with (\ref{Uchi}), hence outside 
of the supergravity SO(8).  Importantly,  this separation would not
work on $\vp^{\bi\bj\bk}$ which is why the remaining two Goldstinos must be removed
before making the above split. Restricting to the diagonal SU(3) subgroup 
(\ref{diag}) we recover the SU(3) assignments in (\ref{Fermions1}).
We will see below that SU(3)$_f$ cannot be maintained as a symmetry
in the presence of electroweak interactions, unlike SU(3)$_c$.
A simultaneous realization of  SU(3)$_f$ and the chiral SM symmetries as 
non-commuting symmetries could be envisaged only in the Planck 
regime with unbroken $\KE$. As the latter does not admit 
a local realisation within space-time based quantum field theory 
this may explain why attempts at a group theoretical family 
unification have not met with success.

The fact that SU(48) can act chirally motivates us to look for a 
realization of {\em all} \, SM symmetries, where the full SM symmetries 
are realized as a subgroup of SU(48), chiral transformations
being those which treat barred and unbarred indices as independent.

\noindent{\bf Electroweak hypercharges and SU(2)$_w\,$:}
The penultimate column in (\ref{Fermions2}) lists the electroweak hypercharges, 
whose implementation entails a further symmetry deformation 
away from the supergravity symmetries, and now also involves
{\em complex} transformations. The hypercharge eigenvalues $y_i$
follow from the standard formula
\beq\label{Y}
Q_{em}   \,=\,  T_3 \,+\,Y
\eeq
where the (hermitean) electroweak isospin operator $T_3$ acts by 
\beq\label{T3}
i T_3 \,=\, \frac14 \Pi_- \otimes 
\Big( \cZ_{1,3} \otimes \bE + \bE \otimes \cZ_{1,3} \Big)
\eeq
on the tri-spinor components (\ref{Fermions2}).
The hermitian projector matrices $\Pi_\pm$ here are defined by
\beq\label{Pi}
\Pi_\pm \,=\, \frac12 \Big( \bE \pm i\cZ_{1,-1} \Big) \,=\, \Pi_\pm^\dagger
\qquad\Ra \quad \Pi_\pm^2 = \Pi_\pm\;,\;\; \Pi_+ +\Pi_- = \bE
\eeq
and take the place of the usual chiral projectors because we are
here working only with left-chiral fields for particles and 
anti-particles. Due to the presence of $i$, these projectors entail 
a complexification, whence any symmetry that acts only on one chirality 
implies complex transformations. Thus (\ref{T3}) realizes 
complex unitary transformations on the original tri-spinor $\chi^{ijk}$.

To extend the weak isospin operator $T_3$ to the full electroweak SU(2)$_w$ symmetry,
we set $T^{\pm} = \frac12(T_1 \pm i T_2)$ and define
\bea\label{Tpm}
T^- (\vp^{\bi|\bbj \bV}) \,&:=&\, \frac12 \eps^{\bbj\bbk\bbl} \vp^{\bi|\bbk\bbl}\;\; , \quad
T^+ (\vp^{\bi|\bbj\bbk}) \,:=\, \eps^{\bbj\bbk \bbl} \vp^{\bi|\bbl\bV}  \nn \\[2mm]
T^- (\vp^{\bbV|\bi \bj}) \,&:=&\, \de^{\bi\bbk} \de^{\bj\bbl} \vp^{\bbV|\bbk\bbl}\;\; , \quad
T^+ (\vp^{\bbV|\bbi\bbj}) \,:=\, \de^{\bbi\bk} \de^{\bbj\bl} \vp^{\bbV|\bk\bl}
\eea
with zero action on the remaining fields, so these operators do not 
act on left-chiral spin-$\frac12$ anti-fermions, {\em viz.}
\beq
T^\pm \,\circ\, (\Pi_+\otimes\bE\otimes \bE) \, = \, 0
\eeq
With these definitions all generators of SU(2)$_w$ act unitarily, and
thus generate transformations in SU(48). It is also evident that
the operations (\ref{T3}) and (\ref{Tpm}) are not compatible with 
SU(3)$_f$, as they mix triplets $\bD_f$ and anti-triplets $\bbD_f$.
The hypothetical family symmetry SU(3)$_f$ is thus broken.

After these preparations we can spell out the association of the 48 spin-$\frac12$ 
fermions (\ref{Fermions2}) with the 48 SM fermions more explicitly,
keeping in mind that this identification could still be subject to a redefinition 
within $\KE$. For the left-chiral electroweak doublets we have
\bea\label{SM1}
\rm{Q}^{\bi \rI}_i \,& \equiv&\,  \left\{
\begin{aligned}
U^{\bfa \rI}{} \equiv  (u^\bfa, c^\bfa ,t^\bfa)_L \,=\, 
\big(\vp^{\bfa |\bbE\bV}, \vp^{\bfa |\bbZ\bV},\vp^{\bfa|\bbD \bV}\big) \\[1mm]
D^{\bfa \rI} \equiv  (d^\bfa, s^\bfa , b^\bfa)_L  \,=\,
\big( \vp^{\bfa|\bbZ\bbD}, \vp^{\bfa|\bbD\bbE},\vp^{\bfa|\bbE\bbZ}\big)\\
\end{aligned}  \right.
\nn\\[2mm]
\rm{L}^\rI_i \,&\equiv &\,    \left\{
\begin{aligned}
N^{\rI\  } \equiv (\nu_e, \nu_\mu , \nu_\tau)_L \,=\,
\big(\vp^{\bbV| \bZ\bD}, \vp^{\bbV|\bD\bE}, \vp^{\bbV|\bE\bZ}\big) \\[1mm]
E^{\rI} \equiv (e^-, \mu^- ,\tau^-)_L \,=\,
\big( \vp^{\bbV|\bbZ\bbD} , \vp^{\bbV|\bbD\bbE} , \vp^{\bbV|\bbE\bbZ} \big)
\end{aligned}  \right.
\eea
For the left-chiral {\em anti}-particles ({\em alias}  right-chiral electroweak singlets) we have
\bea\label{SM2}
\bar U^{\bbfa \rI}{} &\equiv&  (\bar u^{\bbfa}, \bar c^{\bbfa} , \bar t^{\bbfa})_L \equiv 
\big(\vp^{\bbfa |\bE\bbV}, \vp^{\bbfa | \bZ\bbV},\vp^{\bbfa\bD |\bbV}\big) \nn\\[2mm]
\bar D^{\bbfa \rI} &\equiv&  (\bar d^{\bbfa}, \bar s^{\bbfa} , \bar b^{\bbfa})_L  \equiv
\big( \vp^{\bbfa|\bZ\bD}, \vp^{\bbfa|\bD\bE},\vp^{\bbfa|\bE\bZ}\big)\nn\\[2mm]
\bar N^{ \rI} &\equiv& (\bar\nu_e, \bar\nu_\mu , \bar\nu_\tau)_L \equiv
\big(\vp^{\bV|\bbZ\bbD}, \vp^{\bV|\bbD\bbE}, \vp^{\bV|\bbE\bbZ}\big) \nn\\[2mm]
\bar E^{\rI} &\equiv& (e^+, \mu^+ ,\tau^+)_L \equiv
\big( \vp^{\bV|\bZ\bD} , \vp^{\bV|\bD\bE} , \vp^{\bV|\bE\bZ} \big)
\eea
In the absence of a family symmetry the index $\rI=1,2,3$ serves 
merely to label the families, whereas the first index is acted on by SU(3)$_c$. 
The index $\ri = 1,2$ in (\ref{SM1}) refers to SU(2)$_w$.

\noindent{\bf $(B-L)$ symmetry:}
The SM is invariant under another symmetry, U(1)$_{B-L}$, where $B$ and $L$,
respectively, denote baryon and lepton number. This is a global vector-like symmetry,
but since it does not produce chiral anomalies when inserted into triangle diagrams
it can possibly be gauged by means of a $Z'$ vector boson. Such an extension
of the SM gauge group is entirely compatible with the present framework,
as the U(1)$_{B-L}$ symmetry can also be realized via an action on the tri-spinor,
to reproduce the last column of (\ref{Fermions2}). 
The relevant charge operator is again given by a very simple formula
\bea\label{BmL}
\ri Q_{B-L} \,&=&\, 2 \ri Q_{em} -\cJ \nn\\[2mm]
\,&=&\, \frac13 \Big(\cZ_{1,3} \otimes {\bf 1} \otimes {\bf 1} \, + \,  
{\bf 1} \otimes \cZ_{1,3} \otimes {\bf 1} 
               \, + \,  {\bf 1} \otimes {\bf 1} \otimes \cZ_{1,3} \, - \, 2\cJ \Big)
\eea
Because of the presence of the shift $\cJ$ in this formula this operator 
is again outside of the SU(8) $R$-symmetry of $N\eq 8$ supergravity, 
like the electric charge operator.
Because we know the action of $\cJ$ on the gravitinos by (\ref{cJgrav})
the hypercharge eigenvalues of the gravitinos can also be inferred.

\noindent{\bf Gravitino charges\,:} Finally we explain the charges of the 
gravitinos which also follow from the above considerations (after
the absorption of the Goldstinos). For the electric charges we make 
use of (\ref{cJgrav}) to infer that the electric charges of the color singlet 
and color triplet gravitinos, respectively, are given by the fractional 
values $\pm \frac23$ and $\pm \frac13$. The fact that SU(2)$_w$, 
and thus $T_3$ do not act on the gravitinos allows us to deduce 
from (\ref{T3}) that their electroweak hypercharges coincide with their
electric charges. In other words, in addition to strong and electromagnetic
interactions the gravitinos also couple via their electroweak hypercharges,
but in a vector-like way. Similarly, their $(B-L)$ charges can be computed
from (\ref{BmL}), giving $-\frac13$ and $+\frac13$ for the color triplet and
color singlet gravitinos, respectively. These (perhaps counterintuitive)
results do not appreciably affect possible observational consequences,
as the massive gravitinos would manifest themselves mostly through
their electromagnetic interactions \cite{KLMN}, but might affect
baryo- and leptogenesis, preserving $(B-L)$ while producing
matter-antimatter asymmetry, in the sense that non-vanishing $(B-L)$
might be `stored' in gravitinos rather than (anti-)neutrinos. Importantly, 
the gravitinos are thus only subject to vector-like interactions, whence they 
do not contribute to chiral anomalies. Therefore their natural mass
scale is the Planck scale.

With these identifications we have not only fully recovered the correct 
assignments of all SM spin-$\frac12$ fermions at the kinematical level,
but also derived the charge assignments for the gravitinos.
Of course, when expressed in terms of the original tri-spinor $\chi^{ijk}$
the relevant transformations of the spin-$\frac12$ fermions take 
a rather more complicated form.

\section{Embedding SM symmetries into K(E$_{10}$)}

In adapting the original charge assignments of the $N\eq 8$ supergravity
fermions to those of the SM fermions in a `bottom up' procedure we have 
extended the symmetries way beyond the original SU(8) $R$-symmetry of 
this theory. Because this symmetry enhancement is likewise incompatible 
with the original local supersymmetry, the required deformation hints at 
an entirely novel mechanism to eliminate supersymmetry, but in such 
a way that the theory stays `close' to the original supergravity theory. 
It is therefore an important question how to justify such  a departure 
from established field theoretical wisdom. In this section we present 
a possible `top down' justification for this radical step based on the
import of entirely new symmetries.

As already pointed out in the introduction, our main hypothesis 
is that the necessary symmetry enhancement is realized by 
means of the infinite-dimensional $R$-symmetry $\KE$, an 
infinite-dimensional prolongation of the $R$-symmetry SU(8) of
$N\eq 8$ supergravity. $\KE$ is the `maximal compact subgroup' 
\footnote{Here, and in the rest of this article,  we are very cavalier about the 
 distinction between the Lie group and the Lie algebra. The question of 
 how to exponentiate infinite-dimensional, and especially indefinite,  Kac--Moody Lie 
 algebras to actual Lie groups is still very much under debate in 
 the current literature, see \cite{Marquis}.}
of the conjectured duality symmetry $\E$, whose associated 
Lie algebra is defined recursively via the Chevalley-Serre
presentation and the Dynkin diagram of Fig.~1 (see \cite{Kac}
for a general introduction to the theory of Kac-Moody algebras).
The Lie algebra of $\KE$ is then defined as the fixed point `involutory' 
subalgebra of $\E$ left invariant by the Cartan-Chevalley involution. 
Like $\E$ it admits a presentation in terms of generators $x_\ri$ 
(for $\ri,\rj =1,\cdots,10$) and relations \cite{Berman} 
(`Berman-Serre relations')
\begin{align}\label{Berman}
[x_\ri, [x_\ri,x_\rj]] + x_\rj = 0 & \qquad\quad \text{if $\ri,\rj$ are adjacent nodes} , \nn\\[2mm]
[x_\ri,x_\rj] = 0 & \qquad\quad \text{otherwise} .
\end{align}
where $x_\ri = e_\ri - f_\ri$ in terms of the original
Chevalley-Serre generators $e_\ri$ and $f_\ri$. 
The Lie algebra $\KE$ is then defined as the free Lie
algebra over the generators $x_\ri$ modulo these relations.
For further details and explanations we refer to the 
review article \cite{Vigano}.

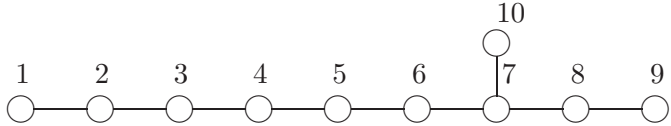
\begin{figure}
\centering
\begin{picture}(200,50)
\put(0,0){\circle{10}}
\put(5,0){\line(1,0){20}}
\put(30,0){\circle{10}}
\put(35,0){\line(1,0){20}}
\put(60,0){\circle{10}}
\put(65,0){\line(1,0){20}}
\put(90,0){\circle{10}}
\put(95,0){\line(1,0){20}}
\put(120,0){\circle{10}}
\put(125,0){\line(1,0){20}}
\put(150,0){\circle{10}}
\put(155,0){\line(1,0){20}}
\put(180,0){\circle{10}}
\put(185,0){\line(1,0){20}}
\put(210,0){\circle{10}}
\put(215,0){\line(1,0){20}}
\put(240,0){\circle{10}}
\put(180,5){\line(0,1){15}}
\put(180,25){\circle{10}}
\put(-2,10){$1$}
\put(28,10){$2$}
\put(58,10){$3$}
\put(88,10){$4$}
\put(118,10){$5$}
\put(148,10){$6$}
\put(182,10){$7$}
\put(208,10){$8$}
\put(238,10){$9$}
\put(180,33){$10$}
\end{picture}\\[4mm]
\caption{Dynkin diagram of $\E$}
\label{dynkin:e}
\end{figure}

Crucial to our considerations is the fact that the hugely infinite-dimensional
Lie algebra defined by (\ref{Berman}) has {\em finite-dimensional} 
unfaithful representations. This is a consequence of the fact that
$\KE$ is {\em not} a Kac--Moody algebra, and thus admits non-trivial
ideals, unlike $\E$, and more generally all Kac-Moody algebras which admit
only the trivial representation as a finite-dimensional representation.
More specifically, analysis of the fermionic sector of $D=11$
supergravity \cite{DKN,dBHP}  has revealed that $\KE$ acts on the 
fermionic states of this theory, and hence also on the $N\eq 8$ 
supermultiplet via such representations, more precisely on the  
fermionic fields {\em at a given spatial point}, see (\ref{spatial}).
So far, only a handful of finite-dimensional unfaithful spinorial 
(double-valued) representations of $\KE$ are known \cite{Vigano}, 
the  two simplest of which are indeed in one-to-one correspondence 
with the spinors of maximal supergravity. Importantly, the results obtained 
extend beyond representation theory and mere kinematics; for instance, as shown 
in \cite{DKN}, the $D=11$ Rarita-Schwinger equation can be reformulated, 
modulo spatial gradients, as a $\KE$-covariant `Dirac equation'.

The key step in arriving at the symmetry enlargement described in section~3 
is the incorporation of $\cJ$ into $\KE$. This follows from the fact that
$\cI\otimes\cI\otimes\cI$  can be realized as an element of $\KE$ acting 
on the 320-dimensional Rarita-Schwinger representation \cite{KN}, and  
hence on all fermionic states of the $N\eq 8$ supermultiplet. 
Consequently,  the action of {\em all} symmetry generators obtained by 
commuting $\cJ$ with SU(8) can be realized as elements of $\KE$,
as we trivially have SU(8)$\,\equiv \, K({\rm E}_7) \subset \KE$.
Furthermore,  this step also allows to define $\cJ$ on the $D=4$ gravitinos,
{\em cf.} (\ref{cJgrav}), a result that cannot be inferred from
the action of $\cJ$ on the spin-$\frac12$ fermions alone \cite{KN}. 
While this extra generator admits a relatively simple realization 
as a $\KE$ element, this is not so for the other SM generators
(\ref{Uc}) and (\ref{Uf}), or (\ref{T3}) and (\ref{Tpm}). Because all
simple $\E$ roots are needed to realize $\cJ$ in $\KE$ \cite{KN}, 
we have to blow up the SM symmetries all the way to a hyperbolic
Kac-Moody group!

The action of $\KE$ on unfaithful representations can  be conveniently 
described in terms of the finite-dimensional quotient groups $\mG_\cR = \KE/\mN_\cR$, 
where $\mN_\cR$ is the `normal subgroup' associated with the 
ideal spanned by those $\KE$ generators that annihilate all elements of 
the given representation space $\cR$ \cite{Vigano}. We caution readers, that -- 
apart from the fact that it is not properly understood how to exponentiate the
$\KE$ generators to a `group' --  $\mN_\cR$ is a rather singular object
 because its associated Lie algebra is of finite co-dimension in $\KE$,
 hence the quotation marks (this point can be somewhat better understood 
 in the affine context \cite{KNP,KKLN,KN2}). Nevertheless the quotient algebra 
 and the quotient groups $\mG$ are perfectly well-defined objects, 
 indicating how the full action of $\KE$ projects onto
 the given finite-dimensional representation. Importantly,
 the quotient algebras are {\em not} Lie subalgebras of $\KE$ 
(as they would be for finite-dimensional Lie algebras). 
 
 For the so-called Dirac representation, of (real) dimension 32 and
 corresponding to the $D=11$ supersymmetry parameter, we have 
\beq
\mG_D \,=\, {\rm SO(32)} \;, 
\eeq
while for the Rarita-Schwinger  representation, of dimension 320, 
and corresponding to the spatial components of the $D=11$ gravitino 
vector-spinor (in a temporal gauge \cite{DKN}) we have \cite{Vigano}
\beq
\mG_{RS} \,=\,  {\rm SO}(288,32)
\eeq
The latter group also contains transformations acting chirally on 
the $N\eq 8$ fermions after reduction to four dimensions. We expect there to 
exist bigger and less unfaithful representations which can possibly capture more 
of the spatial dependence \cite{KN1}. The idea is then that the 
spatial dependence would emerge in a more complete description 
based on such more faithful realizations of $\KE$. Note that these
quotient groups are {\em not} subgroups of $\KE$.

Upon dimensional reduction corresponding to a $3+7$ split
of the spatial coordinates necessary to decompose the
$D=11$ gravitino vector-spinor into eight $D=4$ gravitinos
and 56 spin-$\frac12$ fermions, we arrive at the embeddings
\beq
{\rm SO}(32) \;\supset\;  {\rm U}(2) \times {\rm SU}(8)
\eeq
where SU(2)$\,\subset\,$U(2) corresponds to spatial rotations, and
SU(8) is the $R$-symmetry of $N\eq 8$ supergravity.
For $\mG_{RS}$, we similarly split
\beq
{\rm SO}(288,32) \;\supset\;  {\rm SO}(64,32) \,\times \, {\rm SO}(224) 
\eeq
and then
\bea\label{SU56}
{\rm SO}(64,32) \;&\supset& {\rm U}(4,2) \times {\rm SU}(8) \;\;\nn\\[2mm]
{\rm SO}(224) \;&\supset& \; {\rm U}(2) \times {\rm SU}(56) 
\eea
Here U(2) and U(4,2) are the quotient groups associated, respectively,
to the spin-$\frac12$ and spin-$\frac32$ representations
of K(AE$_3$), the enlarged $R$-symmetry group associated to
pure $D=4$ supergravity \cite{DS} (see also \cite{Vigano}), which contains the 
proper action of the rotation group SU(2) on these fermion 
representations.~\footnote{The algebra $AE_3$ (`Feingold-Frenkel algebra')
 associated to pure Einstein gravity in $D=4$ is a hyperbolic 
 Kac-Moody subalgebra of $\E$, with simple roots $\alpha_1, \alpha_2$ and 
 $\theta \equiv  2\alpha_3 + 3\al_4 + 4\al_5 + 5\al_6 + 6\al_7 + 4\al_8 
 + 2\al_9 + 3\al_{10}$ (the highest E$_8$ root). We thank Axel Kleinschmidt
 for a discussion of this point.}
The SU(56) appearing in the second line of (\ref{SU56}) is just the 
SU(56) of the foregoing section while the SU(8) coincides with the 
$N\eq 8$ $R$-symmetry. Importantly, this step requires 
the full $\E$ symmetry, as the embedding $AE_3\,\times\,$E$_{7} \subset \E$ 
would only give SU(8) as an internal symmetry upon
restricting to the involutory subgroup K($AE_3\,\times\,$E$_7$) =
K($AE_3)\,\times\,$SU(8). We have thus recovered the
relevant representation for the 56 spin-$\frac12$ fermions
as these fermions emerge from the $D=11$ gravitino vector spinor. 
Let us also mention that the extra U(1) factor in both U(2) and U(4,2)
is associated with chiral $\gamma^5$ transformations \cite{DS}, 
transforming dotted and undotted SL(2,${\mathbb{C}}$) in (\ref{Uchi}) 
with opposite phases. 

The unfaithfulness of the fermion representations means that the action 
of the SM symmetries exhibited in the foregoing section is replicated 
by infinitely many $\KE$ generators. This suggests that, with less and less
unfaithful representations capturing more and more of the spatial 
dependence of the supergravity fermions, this infinite replication 
would unfold to a proper gauge group action in the emergent space-time 
effective field theory. Let us emphasize again that the SM groups realized
as proper subgroups of $\mG$ are {\em not} subgroups of $\KE$
because the quotient groups $\mG$ themselves are not subgroups of 
$\KE$ \cite{Vigano}; this feature again points to the necessity of 
identifying larger
unfaithful representations to accommodate proper space-time gauge
symmetries. Clearly, these features are very different from more 
established ways of extracting gauge symmetries from
a Planck scale theory. 

With the bosonic degrees of freedom we have, in a sense, the opposite
problem. Because the bosons transform in a non-linearly realized $\E$
representation \cite{DHN}, and because $\E$ does not have non-trivial 
finite-dimensional representations, the bosons of the $\E/\KE$ coset model 
comprise an infinite number of degrees of freedom, of which only the very 
lowest levels match with the supergravity fields up to and including
first order spatial gradients, whereas the physical interpretation of the higher
level degrees of freedom, whose number increases exponentially 
with the level (see {\em e.g.} \cite{NF}) remains in the dark.

\section{Dynamics?}
All arguments up to this point are entirely kinematical, 
and one may thus ask how to arrive at a possible dynamical 
justification, and to understand why the infinite-dimensional 
symmetry $\KE$ should be broken at the Planck scale in the
way described above, leaving traces of $N\eq 8$ supergravity 
at the lowest energies and giving rise to the SM fermions
in the emergent effective space-time field theory. The dynamics 
of the `spinning particle' $\E/\KE$ sigma model \cite{DHN,DKN} remains
largely unknown: while it provides a perfect match between the $\E$ 
and the (bosonic {\em and} fermionic) field  equations of maximal 
supergravity at the very lowest levels, and modulo higher
order spatial gradients, all attempts at extending 
the known `dictionary' further have been unsuccessful so far. 
Nevertheless, one may still try to exploit the `proximity' of maximal 
supergravity, and the non-trivial facts that (1) four is the preferred
number of space-time dimensions in (compactified) maximal supergravity \cite{FR,DNP},
and (2) the SU(3)$\,\times\,$U(1) stationary point does provide a 
realization of Gell-Mann's scheme, which is our starting point.
Accordingly, in this section we very briefly recall some pertinent facts 
about the stationary structure of gauged maximal supergravities, 
to argue that the SU(3)$\times$U(1) breaking is distinguished.

Like all $N\geq 2$ supergravity theories, gauged $N\eq 8$ supergravity gives rise 
only to vector-like interactions. Accordingly, we here make the hypothesis
that the vector-like interactions in our scheme coincide with those of the 
original theory, albeit in a deformed realization, as they do for the 
SU(3)$\,\times\,$U(1) stationary point. 
For the chiral interactions of the non-Goldstino fermions 
we must assume that the electroweak symmetries and interactions 
are generated dynamically, thus realizing early suggestions in \cite{CJ} 
(see also \cite{EGMZ,EGGZ}), though in a very different way. This step is 
crucial, as the chiral symmetry is what can keep the non-Goldstino fermions 
massless in comparison with the Planck scale, whereas the natural mass
scale for the gravitinos is the Planck scale, thanks to their non-participation
in chiral interactions in our scheme. An important consistency requirement here 
is that whatever chiral subgroup is generated dynamically, it must be 
such that all chiral anomalies cancel, as required for the 
emergent space-time quantum field theories (see {\em e.g.} \cite{Bertlmann}).
As is well known \cite{BIM,GJ} this is an extremely constraining 
condition which in practice fixes the SM charges uniquely. 
Indeed, the hypercharge assignments 
in (\ref{Fermions2}) do meet this requirement, since 
\beq\label{anomaly1}
\sum_{\rm quarks} y_i \,=\, \sum_{\rm leptons} y_i \,=\, 0\;\;, \quad \sum_i \, y_i^3  \,=\, 0
\eeq 
The first two relations also ensure the absence of gravitational anomalies \cite{AGW}. 
Furthermore, the absence of $(B-L)$ induced anomalies follows from
\beq
\sum_i  \, b_i y_i^2 \,=\, \sum_i \, b_i^2 y_i \,=\, 0
\eeq
which is likewise satisfied by (\ref{Fermions2}). Here $y_i$ and $b_i$ 
denote the charge eigenvalues from the last two columns of (\ref{Fermions2}). 
Because the gravitinos participate only in vector-like interactions they do not
contribute to anomalies. Hence no further enlargement of chiral
symmetries beyond SU(2)$\,\times\,$U(1) is possible. See also \cite{Der} 
for an early argument (not directly related to our present considerations), 
according to which the maximal anomaly free chiral subgroup of SU(8) 
that could be dynamically generated within $N\eq 8$ supergravity
is SU(2)$\,\times\,$U(1).

The stationary points of the scalar  field potential of maximal 
SO(8) gauged supergravity discovered first were those 
containing SU(3) as a residual gauge group, with residual gauge
symmetries SO(8), SO(7)$^\pm$, G$_2$, SU(3)$\times$U(1) 
and SU(4), respectively \cite{Warner}. Out of these, only the supersymmetric 
ones with symmetry groups SO(8), G$_2$ and SU(3)$\times$U(1)
are stable (in the sense of satisfying the Breitenlohner-Freedman
condition). Meanwhile, exploiting clever new techniques, many more 
new vacua have been discovered \cite{Fischbacher,DI}. Among these 
there are two more supersymmetric ones, but with very small residual 
symmetries \cite{u1u1,so3}, all others are unstable. 
The unstable vacua being of no interest in the present 
context, we are only left with a very small number of cases to
examine for a possible additional realisation of chiral symmetries on top of 
the residual vector-like gauge symmetries, in such a way that 
all chiral anomalies cancel. If this cannot be done the natural
scale for all fermions is the Planck scale. 

Let us for example  take a look at the G$_2$ stationary point which 
has an $N=1$ residual supersymmetry. Under G$_2$  the eight 
spin-$\frac32$ and 56 spin-$\frac12$ fermions decompose as
\beq
{\bf 8} \,\ra\, \bE\oplus {\bf 7} \;\;,\quad
{\bf 56} \, \ra\, \bE \oplus {\bf 7} \oplus {\bf 7} \oplus {\bf 14} \oplus {\bf 27}
\eeq
After removal of 7+1 Goldstinos we are thus left with 48
spin-$\frac12$ fermions in the representations
 ${\bf 7} \oplus {\bf 14} \oplus {\bf 27}$. Even if one tries to
 somehow associate G$_2$ with strong interactions
(see \cite{GG1,GG2} for very early attempts in this direction), 
 we would then only have one family with these three G$_2$ representations. 
 There is no room for a commuting vector-like U(1),  nor for a chiral SU(2) symmetry.
 Postulating an analog of a chiral U(1)$_Y$ one easily checks that the 
 analog of the cancellation conditions (\ref{anomaly1}) would lead
 to irrational values for the hypercharges, which is not acceptable.
 For these reasons we conclude that the G$_2$ vacuum is not amenable to a
 deformation procedure with subsequent chiral symmetry enhancement.
 
There are other gaugings of  $N\eq 8$ supergravity which also admit
an SU(3)$\,\times\,$U(1) vacuum with $N\eq 2$ supersymmetry. This is the
case for the $\omega$-deformed theories of \cite{omega,inverso,Fisch1} and
the ISO(7) gauging of \cite{GJV,YPV}. While the $\omega$-deformed 
theories have no known higher-dimensional origin, the ISO(7) gauging
originates from massive IIA supergravity \cite{YPV} (the mIIA theory also 
fits into the $\E$ framework \cite{mIIA}). These vacua are related, 
in the sense that the corresponding gauged theories
differ only by the choice of embedding tensor which transforms
in the ${\bf 912}$ of E$_{7}$; for instance, the spectra of \cite{NW} and 
\cite{YPV} are related to each other by an SO(8) triality rotation.
The charges and representations arising in the decomposition of the 
$N\eq 8$ supermultiplet are completely fixed by $N\eq 2$ supersymmetry \cite{NW}.
Hence the deformation and symmetry enlargement procedure giving
(\ref{Fermions2}) continue to apply.

\section{Outlook}

There remain many open problems. The ones that arise in connection
with the M theory proposal of \cite{DHN,DKN} have already been discussed
at length in the literature, such as the question of how to incorporate higher order
spatial gradients and to explain in detail the emergence of space-time 
geometry from the `spinning' $\E$ sigma model or some extension of it, or the 
problem of understanding the physical significance of the higher
level degrees of freedom in $\E$. Likewise, the construction of a tower
of less and less unfaithful fermionic representations of $\KE$ 
(for the much simpler case of K(E$_9$) partial progress on this front has 
been achieved in \cite{KKLN}) to capture more and more of the spatial 
dependence of the supergravity fermions remains an open problem. 
It is clear that this will at the very least require a much better understanding of the mathematics
of hyperbolic Kac-Moody algebras (and groups) than currently available.

However, the present work raises specific issues motivated by the desire 
to establish a contact with SM physics, at least as far as the fermionic 
sectors are concerned. These are:

\begin{itemize}         
\item Can we understand $\E$ and $\KE$ dynamics in order to explain 
         the symmetry breaking that would lead to the implementation 
         of SM symmetries on the SM fermions?
\item We have exhibited the `pointlike' action of the SM symmetries
         on the fermions (\ref{spatial}), but at this point we cannot tell the 
         difference between a global (rigid) and a local (gauge) symmetry, 
         as this would require the proper incorporation of higher order spatial gradients. 
         The SM symmetries being local, this would constrain their action
         on less unfaithful representations of $\KE$ in a very specific way.
\item In what sense precisely are the $\E$ framework and $N\eq 8$ supergravity 
         `close' to one another, and what do $\E$ and $\KE$ `know' about the 
         supergravity stationary points? This is again a question that probably cannot 
         be answered without incorporating full spatial dependence.
\item How can we explain the dynamical generation of chiral symmetries 
         and interactions in the emergent effective field theory? The idea that electroweak 
         interactions could involve some degree of compositeness is not new, but
         no working model for such a scenario has ever been put forward.
\item What is the ultimate fate of space-time supersymmetry? Is it broken by a novel
         mechanism at the Planck scale, or is it absent altogether in any form in the
         bigger framework of the present paper, in the sense that $\E$ symmetry
         supersedes supersymmetry? Then what is the reason that the fermion 
         representation content and the charge assignments are related to the
         $N\eq 8$ supermultiplet in the way described above?
\item Can we detect traces of supermassive gravitinos in upcoming underground experiments,
         and if so, is there a way to confirm that they carry spin $s=\frac32$ and fractional charges?      
\end{itemize}

\noindent
 {\bf Acknowledgments:} K.A.~Meissner thanks AEI for hospitality and support. 
 We are most grateful to T.~Fischbacher and A.~Kleinschmidt for illuminating 
 discussions and comments on a first version of this paper. We also
 thank O. Varela for discussions and correspondence.

\end{document}